\documentclass[twocolumn,aps]{revtex4}
\usepackage{times}
\usepackage{natbib}
\usepackage{graphicx}
\usepackage{amssymb,amsmath,amsbsy}
\usepackage[usenames]{color}
\definecolor{hellgrau}{gray}{0.6}
\begin{document}
\title{Scheme for generating a sequence of single photons of alternating polarisation}
\author{T. Wilk}
\author{H.\,P. Specht}
\author{S.\,C. Webster}
\author{G. Rempe}
\author{A. Kuhn}
\email[Corresponding author: ]{gerhard.rempe@mpq.mpg.de}
\affiliation{Max-Planck-Institut f\"ur Quantenoptik, Hans-Kopfermann-Str. 1, D-85748 Garching, Germany}
\date{\today}
\begin{abstract}
Single-photons of well-defined polarisation that are
deterministically generated in a single spatio-temporal field mode
are the key to the creation of multi-partite entangled states in
photonic networks. Here, we present a novel scheme to produce such
photons from a single atom in an optical cavity, by means of
vacuum-stimulated Raman transitions between the Zeeman substates of
a single hyperfine state. Upon each transition, a photon is emitted
into the cavity, with a polarisation that depends on the direction
of the Raman process.
\end{abstract}
\maketitle

\section{Introduction}
Worldwide, large efforts are being made to interface single atoms
with single photons, since this is the key to the entanglement
\cite{Cabrillo99, Browne03, Duan03, Blinov04} and teleportation
\cite{Bose99, Lloyd01} of quantum states between distant atoms.
Transform-limited and mutually coherent single photons in
well-defined polarisation and spatio-temporal field modes are needed
so that the desired quantum-correlation effects occur
\cite{Knill01,Hong87,Legero03}. Some crucial problems that arise in
the production of such photons have been solved, such as the choice
of an adequate emitter, the efficient collection of the emitted
photons or their channelling into a given direction. In particular,
several research groups have demonstrated that strongly coupled
atom-cavity systems can be used as deterministic sources of single
photons \cite{Kuhn02,Keller04,McKeever04}. In these experiments, a
single atom located in the cavity is exposed to laser pulses which
drive one branch of a Raman transition. The other branch is resonant
with the cavity mode so that the field mode of the cavity receives
one photon upon each successful Raman transition. The photons are
finally emitted from the cavity, and Hanbury Brown \& Twiss
photon-correlation measurements are used to demonstrate the
single-photon nature of the emission.

\begin{figure*}
\includegraphics[width=\textwidth]{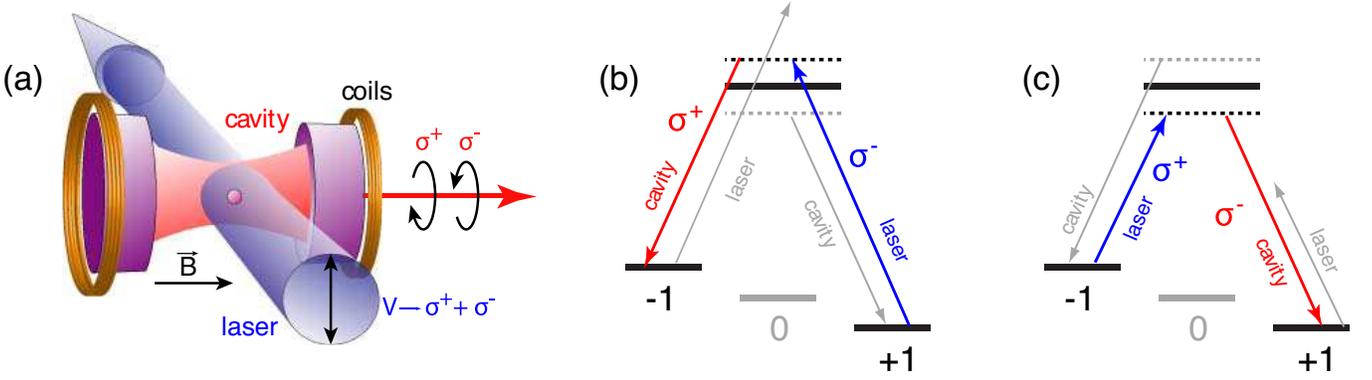}
\caption{Principle of the photon generation scheme. A single atom is
strongly coupled to an optical cavity, and a magnetic field acting
along the cavity axis lifts the degeneracy of the Zeeman substates
(a). From the side, laser pulses drive Raman transitions in the
atom. The polarisation of the laser is perpendicular to the cavity
and B-field axis, so that it decomposes into $\sigma^+$ and
$\sigma^-$ components with respect to the cavity. As illustrated in
(b) and (c), the cavity is resonant with the non-shifted atomic
transition, but the laser pulse is red or blue detuned: a blue
detuned laser pulse drives a Raman transition from $m_F =+1$ to $m_F
=-1$ and generates a $\sigma^+$ polarised photon (b), while a red
detuned laser pulse drives the transition in the opposite direction
and generates a $\sigma^-$ polarised photon (c).}
\label{Fig:1-principle}
\end{figure*}

In this article, we propose to extend this photon-generation scheme
in such a way that single photons of well-defined polarisation are
emitted from a coupled atom-cavity system. This new method is based
on our previous work \cite{Hennrich00,Kuhn02,Hennrich03,Legero04},
where Raman transitions between  hyperfine states of a $^{85}$Rb
atom were used to generate single photons. The large number of
accessible magnetic sublevels in $^{85}$Rb meant that the
polarisation of these photons was undefined. In order to achieve
polarisation control, we now investigate the situation shown in
Fig.\,\ref{Fig:1-principle}. Vacuum-stimulated Raman transitions are
considered between the $m_F =\pm 1$ Zeeman substates of an $F=1$
hyperfine state, e.g. in the electronic ground state of $^{87}$Rb
atoms. We assume a magnetic field along the cavity axis that lifts
the degeneracy of the magnetic substates. If the frequency of the
applied laser pulses is either red or blue detuned from the
unperturbed transition by twice the B-field induced Zeeman shift,
while the cavity frequency is in resonance with the unperturbed
atomic transition, then the transition between the $m_F=+1$ and
$m_F=-1$ levels is resonantly driven by laser and cavity, but with
their roles changing as a function of the chosen laser detuning. In
this way the Raman transition can be driven in one or the other
direction, leading to an emission of either $\sigma^+$ or $\sigma^-$
photons. As can be seen in the level scheme shown in
Fig.\,\ref{Fig:1-principle}, this method requires the laser to be
polarised perpendicular to the cavity axis, so that it decomposes
into $\sigma^+$ and $\sigma^-$ polarisation components with respect
to the magnetic field direction. Only the $\sigma^+$ polarisation
component of the driving laser is used for the generation of a
$\sigma^-$ polarised photon and vice versa. The other polarisation
component of the laser pulse is always present, but it is
out-of-resonance with all relevant atomic transitions.

As the cavity supports both polarisation modes, alternating the
laser frequency between the two possible resonances from pulse to
pulse should be sufficient to generate a sequence of single photons
of alternating polarisation. No repumping of the atom to its initial
state is required from one photon to the next, since the final state
reached with a $\sigma^+$ photon emission is at the same time the
initial state for producing a $\sigma^-$ photon with the subsequent
driving laser pulse, and vice versa. With this novel scheme, one
should therefore be able to produce sequences of photons with
well-defined polarisation.

\section{The ideal three-level atom}
For an illustration of the photon generation process, we first
consider an idealised atom with an $F=1$ hyperfine ground state and
an $F'=0$ electronically excited state. The Zeeman shift induced by
the magnetic field on the $m_F=\pm 1$ atomic ground states is
$\mp\Delta_B=\mp |g_L|\mu_B B$, where $\mu_B$ is the Bohr magneton
and $g_L$ the Land\'{e} factor. In the following the
$m_F=\{-1,0,+1\}$ Zeeman substates of the $F=1$ ground state will be
written as $| {\scriptstyle -}\rangle$,$|0\rangle$ and
$|{\scriptstyle +}\rangle$, whereas the $F'=0$ excited state is
labelled $|e\rangle$. For geometric reasons the cavity supports only
$\sigma^+$ and $\sigma^-$ photons, which have identical frequencies
if the relevant cavity modes are degenerate. The state of the
coupled atom-cavity system can therefore be written as a
superposition of product states $|i,n_+,n_-\rangle$, with $i$
representing the atomic state, and $n_{\pm}$ denoting the number of
$\sigma^{\pm}$ photons. In this paper we restrict the number of
photons in each mode to zero or one, since higher photon numbers are
very unlikely. Note that the cavity couples only product states of
different photon numbers, while the interaction of the atom with the
pump laser leaves the intra-cavity photon number unchanged. The pump
frequency $\omega_p$ and the cavity resonance frequency $\omega_c$
are both close to the $|0\rangle$ to $|e\rangle$ transition
frequency $\omega_{0e}$. We also disregard the states
$|0,n_+,n_-\rangle$, since $|0\rangle$ is decoupled from all other
internal states. In this way an atomic three-level system in a
$\Lambda$-configuration is obtained. Here, we choose the energy of
the excited state $|e,0,0\rangle$ to define the origin of the energy
scale and we divide the Hamiltonian of the system into two parts,
$\hat{H} = \hat{H}_{stat} + \hat{H}_{int}$. The stationary part
$\hat{H}_{stat}$ includes the energy levels of atom and cavity, and
$\hat{H}_{int}$ describes the interaction of the atom with pump
laser and cavity. The system is examined in the interaction picture.
In the rotating wave approximation, the stationary part of the
Hamiltonian reads
\begin{widetext}
\begin{equation}
\hat{H}_{stat} = \hat{H}_{atom} + \hat{H}_{cavity} = \hbar \Big[
\left( \Delta_p + \Delta_B \right)|{\scriptstyle -}\rangle\langle
{\scriptstyle -}| + \left(\Delta_p - \Delta_B \right)|{\scriptstyle
+} \rangle\langle {\scriptstyle +}| \Big] +
\hbar\Delta_{cp}\left(\hat{a}^{\dag}\hat{a}+
\hat{b}^{\dag}\hat{b}\right),\label{eq:Hstat}
\end{equation}
\end{widetext}
where $\Delta_{p} \equiv \omega_p-\omega_{0e}$ is the detuning of
the pump laser from the transition between $|0\rangle$ and
$|e\rangle$ and $\Delta_{cp} \equiv \omega_{c}-\omega_p$ is the
difference between cavity resonance and pump laser frequency. Here,
$\hat{a}^{\dag}$ and $\hat{a}$, or $\hat{b}^{\dag}$ and $\hat{b}$,
are the creation and annihilation operators of a photon in the
$\sigma^+$ or $\sigma^-$ polarised cavity mode, respectively. The
interaction between atom and cavity is given by the interaction
Hamiltonian
\begin{widetext}
\begin{equation}
H_{int} = -\hbar \left[ g\left(|e\rangle\langle {\scriptstyle
-}|\hat{a} + \hat{a}^{\dag}| {\scriptstyle -} \rangle\langle e| +
|e\rangle\langle {\scriptstyle  +}|\hat{b} +
\hat{b}^{\dag}|{\scriptstyle +}\rangle\langle e |\right) +
\frac{1}{2}\Omega\Big(|e\rangle\langle {\scriptstyle -} | +
|{\scriptstyle -}\rangle\langle e| + |e\rangle\langle {\scriptstyle
+} | + | {\scriptstyle  +}\rangle\langle e|\Big)\right],
\label{eq:Hint}
\end{equation}
\end{widetext}
where g is the coupling constant of the atom to both cavity modes
(assumed to be equal), and $\Omega$ is the Rabi frequency of the
pump laser.

The cavity decay gives rise to the emission of photons from the
cavity, which is a non-unitary process. It cannot be included in a
Hermitian Hamiltonian, but its effects on the density matrix can be
expressed by the Liouville operator \cite{Scully97},
\begin{equation}
\hat{L}[\hat{\rho}]= \kappa \left(2\hat a \hat{\rho}\hat a^{\dag}
- \hat a^{\dag}\hat a \hat{\rho} - \hat{\rho}\hat a^{\dag}\hat a +
2\hat b \hat{\rho}\hat b^{\dag} - \hat b^{\dag}\hat b \hat{\rho} -
\hat{\rho}\hat b^{\dag}\hat b \right) .\label{liouville}
\end{equation}
The cavity field decay rate $\kappa$, here identical for both
polarisations, has to be fast with respect to the Raman process to
ensure that the photon is emitted from the cavity before being
reabsorbed by the atom.

The time evolution of the system is then given by the master
equation
\begin{equation}
\frac{d}{dt} \hat{\rho} = -\frac{i}{\hbar} [\hat H, \hat{\rho} ]
+\hat{L}[\hat{\rho}] .\label{eq:masterEQ}
\end{equation}
Note that a similar Raman process driven in only one direction has
been discussed previously in \cite{Kuhn99}. Here, we search for
conditions where Raman transitions can be driven in both directions,
albeit with the cavity frequency fixed. This imposes new
constraints, and we now have to determine the optimal parameters for
efficient production of $\sigma^+$ and $\sigma^-$ photons.

\begin{figure*}
\includegraphics[width=\textwidth]{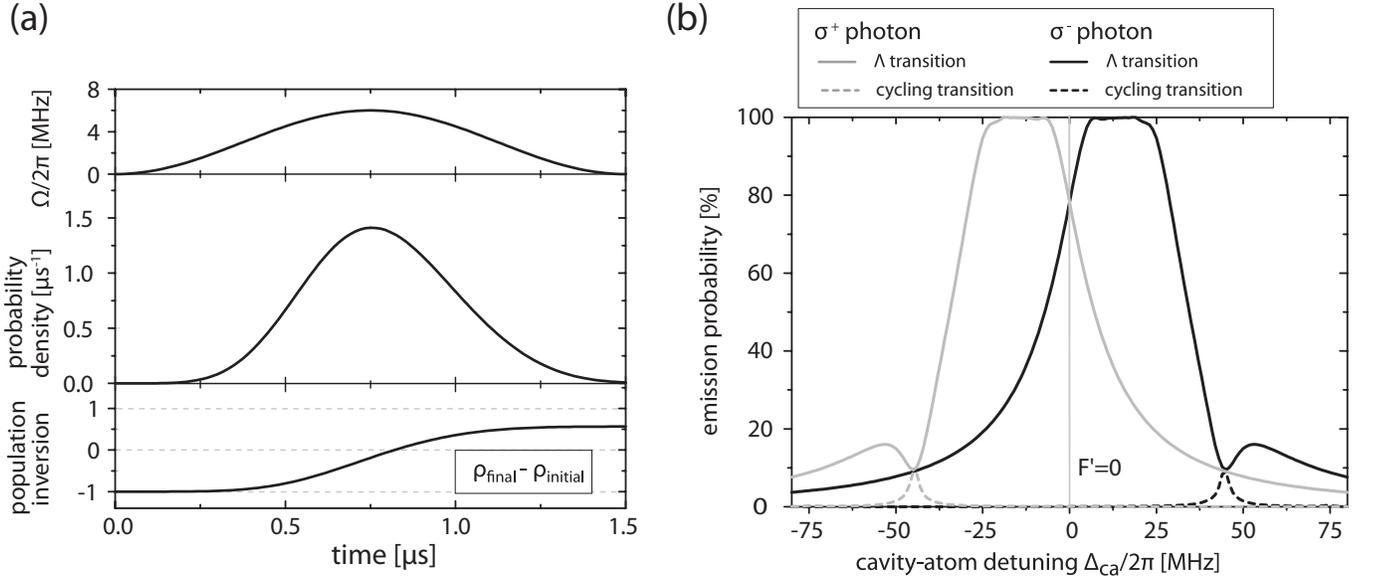}
\caption{Results for the idealised three-level atom with parameters
$(g;\kappa;\Delta_B)/2\pi$ = (2.7; 1.25; 15)\,MHz. \textbf{(a)} Time
evolution of the system: pump laser Rabi-frequency $\Omega/2\pi$
follows a $\sin^2(\pi t/1.5\mu s)$-pulse amplitude, with a peak
value of 6\,MHz. The probability density for generating a photon per
$\mu s$ is shown. Integrating over the whole pulse gives an emission
probability of $78\%$ per pulse. Accordingly, the population
transfer between the atomic state, defined by the difference in
population of the final and initial state, ends up at an inversion
of 0.56. \textbf{(b)} Photon production efficiency as a function of
the cavity-atom detuning $\Delta_{ca}$. The pump laser is always
tuned to the desired Raman resonance: for $\sigma^-$ photons
$\Delta_{cp}= -2\Delta_B$, for $\sigma^+$ photons
$\Delta_{cp}=+\Delta_B$. Black lines stand for the probability for
generating $\sigma^-$ photons, grey lines for $\sigma^+$ photons,
whereas solid lines indicate the desired $\Lambda$-type transition,
and dashed lines the cycling transitions, starting and ending in the
same state.} \label{fig:2-ideal}
\end{figure*}

The initial state of the system is set to
$|\Psi_{start}\rangle=|{\scriptstyle +},0,0\rangle$ and realistic
cavity parameters are chosen,
$(g;\kappa;\Delta_B)/2\pi=(2.7;1.25;15)$\,MHz (e.g. corresponding to
a $^{87}$Rb atom in the 5S$_{1/2}|F=1, m_F=+1\rangle$ hyperfine
ground state and excitation of the D$_2$-line, inside a cavity used
in former experiments \cite{Kuhn02, Hennrich00, Hennrich03,
Legero04} and a magnetic field of 21.4\,G). The cavity is in
resonance with the $|0\rangle$ to $|e\rangle$ transition, (i.e.
$\omega_c=\omega_{0e}$), and in order to resonantly drive the Raman
transition from $|{\scriptstyle +},0,0\rangle$ to $|{\scriptstyle
-},1,0\rangle$, the pump laser has to have a detuning of
$\Delta_p=+2\Delta_B$. The pump-laser Rabi frequency follows a
$\sin^2(\pi t / 1.5\mu s)$-pulse amplitude with a peak value of
$\Omega /2\pi= 6$\,MHz for each polarisation, as shown in
Fig.\,\ref{fig:2-ideal}\,(a).
The probability density for emitting a photon varies as a function
of time, and reflects the envelope of the generated single-photon
wave packet. Note that the shape of the photon depends on the
driving field. Changing the shape of the pump pulse or its peak
value has a direct impact on the emission probability. The photon
can therefore be shaped in many desired ways. The integral of the
probability density over the whole photon duration gives the overall
emission probability of a photon, here it measures $78\%$. The lower
part of Fig.\,\ref{fig:2-ideal}\,(a) illustrates the population
inversion between the states $|{\scriptstyle -}\rangle$ and $|{\scriptstyle +}\rangle$, which is here the population difference between the final and the initial atomic state. The
inversion starts from $-1$ at $t=0$ and increases until it reaches
$+0.56$. Note that any losses out of the system (except for cavity
decay) have been omitted in this simplified model, therefore
population which is not transferred into the other ground state
simply stays in the initial state. Under these conditions, the emission probability equals the fraction of transferred population. This is not the case
in general, since other loss channels than the cavity might exist.

During the population transfer from the state $|{\scriptstyle +}
,0,0\rangle$ to $|{\scriptstyle -},1,0\rangle$, the single photon
state of the cavity decays. Since the time constant 2$\kappa$ is
much faster than the duration of the pump pulse, the photon leaks
out of the cavity during its generation. Once the atom has reached
the state $|{\scriptstyle -},0,0\rangle$, it will be very unlikely
to undergo another Raman transition back to the initial state, since
the back transition is detuned by $4\Delta_B$ (see
Fig.\,\ref{Fig:1-principle} (b), grey lines) which is much larger
than the cavity linewidth of 2$\kappa$. This means a second emission
is suppressed and thus only a single $\sigma^+$ or $\sigma^-$ photon
is generated. The efficiency of the photon generation depends on
many parameters. Here it is $78\%$, but it can rise up to $100\%$
with increasing pump power or with a stronger atom-cavity coupling.
For a larger Zeeman splitting the emission probability would
decrease.

Exactly the same photon generation efficiency and photon envelope
are obtained when the initial state is $|{\scriptstyle
-},0,0\rangle$ and the detuning of the pump laser frequency is
$\Delta_p=-2\Delta_B$. Then Raman resonance is fulfilled for the
$\Lambda$-type transition shown in Fig.\,\ref{Fig:1-principle} (c)
and a $\sigma^-$ photon is generated. Both cases are similar, since
the cavity frequency has been chosen to be on the unshifted
transition from $F=1$ to $F'=0$, leading to an atom-cavity detuning
of $\pm\Delta_B$ with respect to the atomic resonances from
$|{\scriptstyle +}\rangle$ to $|e\rangle$ and $|{\scriptstyle
-}\rangle$ to $|e\rangle$. This symmetry is only broken when we
detune the cavity. In Fig.\,\ref{fig:2-ideal}\,(b), the photon
generation efficiencies are plotted as a function of the cavity-atom
detuning, $\Delta_{ca}\equiv\omega_c -\omega_{0e}$, with the pump
frequency always chosen in such a way that laser and cavity are in
Raman resonance with either the $|{\scriptstyle +}\rangle$ to
$|{\scriptstyle -}\rangle$ or the $|{\scriptstyle -}\rangle$ to
$|{\scriptstyle +}\rangle$ transition, i.e. $\Delta_{cp}=\pm
2\Delta_B$. Black lines indicate the probability to emit a
$\sigma^-$ photon and grey lines a $\sigma^+$ photon. The solid
lines give the efficiencies for the photon production within the
desired $\Lambda$-type Raman transitions, i.e. starting in
$|{\scriptstyle +}\rangle$ and ending in $|{\scriptstyle -}\rangle$
for $\sigma^+$ or vice versa for $\sigma^-$ photons. Driving a
$\Lambda$-type transition with a cavity-atom detuning around
$\Delta_{ca}=-\Delta_B$ for $\sigma^+$ photons (grey line) or
$\Delta_{ca}=+\Delta_B$ for $\sigma^-$ photons (black line) leads to
100$\%$ efficiency of the process. In both cases, the detuning of
the cavity compensates the Zeeman shift and both arms of the Raman
resonance coincide with an atomic resonance. However, to generate
$\sigma^+$ and $\sigma^-$ polarised photons with the same
probability in case of a fixed cavity detuning, we have to accept a
compromise, i.e. $\Delta_{ca}=0$ with an emission probability
reduced to $78\%$.

To verify whether no unwanted second photon will emerge from the
system, we also have calculated the probability for a photon
production from an atom starting in the wrong initial state, e.g.
state $|{\scriptstyle -}\rangle$ for $\sigma^+$ photons. If
$\sigma^+$ photons can be generated in this case, the atom would
undergo a cycling transition, since initial and final state are the
same. As indicated by the grey dashed line (black dashed line for
the analogous case with $\sigma^-$ photons) in
Fig.\,\ref{fig:2-ideal}\,(b) the probability for such cycling
transitions is close to zero in the frequency range around
$\Delta_{ca}=0$, where the scheme for generating photons of
alternating polarisation works best, so no second photon will be
emitted. Only when the pump laser accidently hits an atomic
resonance does the emission probability become non-negligible. This
phenomenon is however an artefact of our simplified model, the peaks
appearing because the spontaneous decay of the excited atomic state
has been omitted. Consequently, once the atom is in the excited
state it can only emit into the far-off resonant cavity mode. In a
real atom the spontaneous decay to other states would dominate the
atom's behaviour. This is discussed in more detail below.

This simple model shows that, for $\Delta_{ca}=0$, single-photon
polarisation control can be achieved by choosing the appropriate
pump-laser frequencies. There is no need to alter the cavity
frequency or the pump polarisation. Moreover, the probability to
emit two photons during the same pump pulse is vanishingly small.

\section{Polarised photons from alkali atoms}
To analyse the behaviour of a coupled atom-cavity system under more
realistic conditions, all relevant atomic levels and the spontaneous
decay of the excited states must be taken into account. As an
example we consider the D$_2$-line of $^{87}$Rb. Its 5S$_{1/2}$
ground state decomposes into two hyperfine states, $F=\{1;2\}$, with
$2\pi\times$6.8\,GHz hyperfine splitting, while the 5P$_{1/2}$
excited state has four hyperfine substates, $F'=\{0;1;2;3\}$, with
splittings of $2\pi\times$(72; 157; 267)\,MHz. The two 5S$_{1/2}
$($F=1$,$m_F=\pm 1$) states are the $|\pm\rangle$ states in our
scheme, while the virtual excited level of the Raman transition is
some superposition of 5P$_{1/2}$($m_{F'}$=0) states. Note that the
origin of our energy scale is chosen to coincide with $F'=0$. This
state is still labelled $|e\rangle$. In the ground state, $F=2$ is
so far from resonance with pump and cavity that there is effectively
no coupling. Therefore spontaneous emissions into $F=2$ constitute
an additional loss channel, but the state as such need not be
considered. Moreover, we restrict ourselves to a situation where the
cavity frequency is near resonant with the transitions from $F=1$ to
$F'=\{0;1\}$. With a distance of $2\pi\times$157\,MHz between $F'=1$
and $F'=2$, the latter state is far from resonance. Combined with
the fact that the dipole-matrix elements for transitions between
$F=1$ and $F'=2$ are smaller than those for the relevant transitions
to $F'=\{0;1\}$ the $F'=2$ state can be neglected. We also omit the
$F'=3$ state, since it does not couple to $F=1$.

\begin{table*}
 \begin{tabular}{@{}ccccc}\toprule
   $\mathcal A_{ij}$ & $|F'=0,m_{F'}=0\rangle$ & $|F'=1,m_{F'}=+1\rangle$ &
   $|F'=1,m_{F'}=0\rangle$ & $|F'=1,m_{F'}=-1\rangle$\\ \colrule
  $|F=1,m_F=+1\rangle$ & $\sqrt{\frac{1}{6}}$ & \textcolor{hellgrau}{$\sqrt{\frac{5}{24}}$} & $-\sqrt{\frac{5}{24}}$ & -
  \\
  $|F=1,m_F=0\rangle$ & \textcolor{hellgrau}{$\sqrt{\frac{1}{6}}$} & $\sqrt{\frac{5}{24}}$ & - &
  $-\sqrt{\frac{5}{24}}$\\
  $|F=1,m_F=-1\rangle$ & $\sqrt{\frac{1}{6}}$ & - & $\sqrt{\frac{5}{24}}$ &
  \textcolor{hellgrau}{$-\sqrt{\frac{5}{24}}$} \\  \botrule
 \end{tabular}
 
 \caption{The angular part of the dipole matrix elements $\mathcal
A_{ij}$ for the transition between $|i\rangle$ and $|j\rangle$. With
these numbers, the Rabi frequencies are $\Omega_{ij} = \mathcal
A_{ij}\cdot\Omega_0$. For a transition where $\Delta m=+1$ the
cavity coupling constant reads $g^+_{ij}=\mathcal A _{ij}\cdot g_0$,
for transitions with $\Delta m=-1$ it is $g^-_{ij}=\mathcal A
_{ij}\cdot g_0$ and $g^{\pm}_{ij}=0$ in all other cases. The indices
$i$ and $j$ refer to the states $|F,m_F\rangle$ and
$|F',m_{F'}\rangle$. The electronic parts of the coupling constant
and Rabi frequency are chosen to be $g_0/2\pi = 6.7$\,MHz and
$\Omega_0/2\pi = 14.7$\,MHz. Note that in our scheme only
$\sigma^{\pm}$ transitions are accessible. The $\mathcal A_{ij}$ for
the non-relevant $\pi$ transitions are shown in grey.}
\label{tab:Aij}
\end{table*}

The stationary part of the Hamiltonian now includes all involved
atomic levels. It reads
\begin{equation}
\hat H_{stat}=\hbar\left( \sum_i\Delta_i |i\rangle\langle i|+
\Delta_{cp}\left(\hat a^{\dag}\hat a+\hat b^{\dag}\hat
b\right)\right).\label{eq:Hstat_all}
\end{equation}
Here, $\Delta_i$ stands for the energy of the respective atomic
level in the rotating frame, including pump detuning and Zeeman
shift with respect to the zero level of our calculation. The
interaction part of the Hamiltonian can be written as
\begin{widetext}
\begin{equation}
\hat H_{int}= - \hbar \sum_{i,j}
\left[g^+_{ij}\left(|i\rangle\langle j|\hat a^{\dag} +
|j\rangle\langle i|\hat a \right)+g^-_{ij}\left(|i\rangle\langle
j|\hat b^{\dag} + |j\rangle\langle i|\hat b \right) +
\frac{1}{2}\Omega_{ij}\Big(|i\rangle\langle j| + |j\rangle\langle
i|\Big)\right],\label{eq:Hint_all}
\end{equation}
\end{widetext}
where $i$ and $j$ denote ground and excited states, respectively.
Transitions between $|i\rangle$ and $|j\rangle$ are either driven by
the pump with a Rabi frequency $\Omega_{ij}$, or by the coupling of
the atom to the $\sigma^{\pm}$ cavity modes, with the atom-cavity
coupling constants $g^{\pm}_{ij}$. Both $\Omega_{ij}$ and
$g^{\pm}_{ij}$ depend on the angular part $\mathcal A_{ij}$ of the
dipole matrix elements listed in
Tab.\,\ref{tab:Aij} \cite{Metcalf99, Steck01}.
We have to distinguish between $\sigma^+$ and $\sigma^-$ cavity
coupling constants.
They read $g^{\pm}_{ij}=\mathcal A _{ij}\cdot g_0$, but 
 $g^+_{ij}$ is zero unless $\Delta m \equiv
m_{F'} - m_F = +1$ and $g^-_{ij}$ is zero unless $\Delta m=-1$. 
For the relevant transitions, also the Rabi frequencies read
$\Omega_{ij} = \mathcal A_{ij}\cdot \Omega_0$. For our calculation,
we have chosen realistic values for the electronic parts of the
atom-cavity coupling and of the peak Rabi frequency, i.e. $g_0/2\pi
= 6.7$\,MHz and $\Omega_0/2\pi = 14.7$\,MHz.

To include the spontaneous decay of the involved excited states, we
extend the Liouville operator (\ref{liouville}) to
\begin{widetext}
\begin{equation}
\hat{L}[\hat{\rho}]=\sum_{i,j}
\left[\gamma_{ij}\Big(2|i\rangle\langle j|\hat{\rho} |j\rangle
\langle i|\Big)\right] - \sum_{j}\gamma_j\Big(|j\rangle\langle
j|\hat{\rho}+\hat{\rho}|j\rangle\langle j|\Big)
\label{liouville_loss} + \kappa \Big(2 \hat a \hat{\rho}\hat
a^{\dag} - \hat a^{\dag}\hat a \hat{\rho} - \hat{\rho}\hat
a^{\dag}\hat a + 2\hat b \hat{\rho}\hat b^{\dag} - \hat b^{\dag}\hat
b \hat{\rho} - \hat{\rho}\hat b^{\dag}\hat b\Big) \nonumber ,
\end{equation}
\end{widetext}
where $\gamma_{ij}$ is the transition strength of the decay channel
from $|j\rangle$ to $|i\rangle$ and $\gamma_j$ the total
polarisation decay rate of the excited state $|j\rangle$, including
transitions to other levels, such as the $F=2$ ground state.

The extension of the model makes the prediction of the emission
probability more difficult. The influence of certain atomic bare
states increases or decreases depending on their distance from the
virtual excited level of the Raman transition. Only from a numerical
simulation of the scheme, using the master equation
(\ref{eq:masterEQ}) with the extended Hamiltonian and Liouvillian
from Eq.\,(\ref{eq:Hstat_all}-\ref{liouville_loss}), do we gain more
insight into the physical processes.

\begin{figure*}
\centerline{\includegraphics[width=\textwidth]{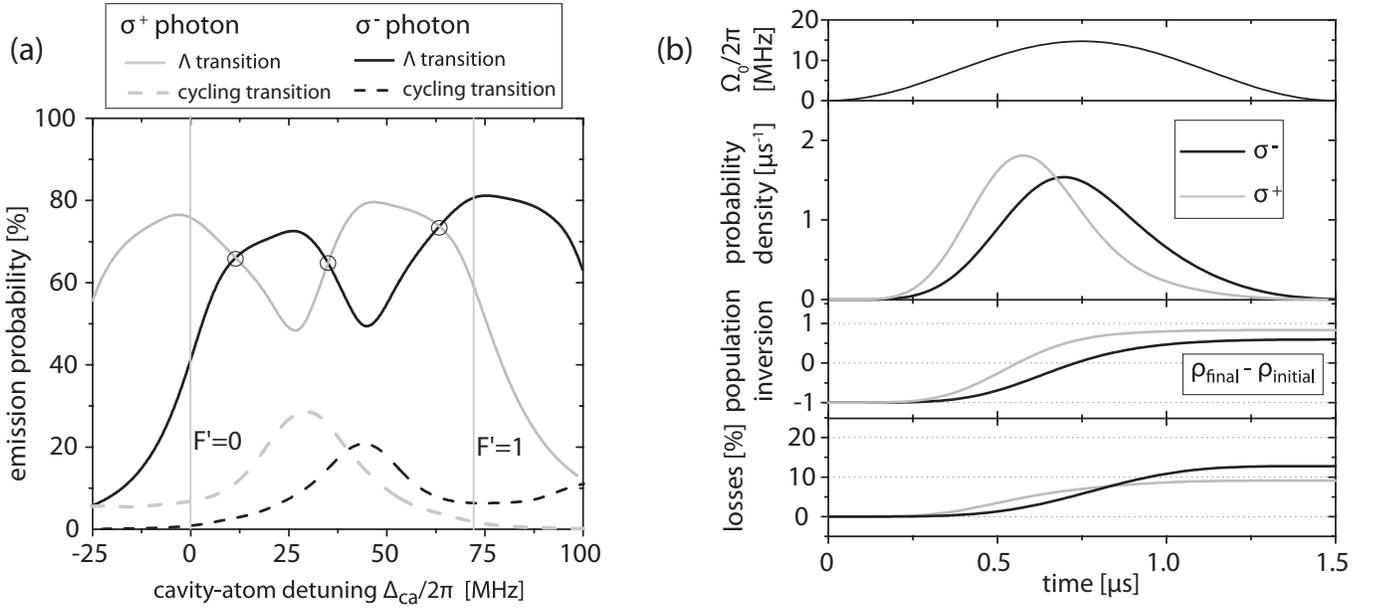}}
\caption{Polarised photons from a $^{87}$Rb atom: (a) Photon
emission probability as a function of the cavity-atom detuning
$\Delta_{ca}$ with respect to the transition from $|0\rangle$ to
$|e\rangle$. The ground state $F=1$ and the excited states $F'=0$
and $F'=1$ are incorporated in the simulation, as well as their
spontaneous decay. The vertical grey lines indicate zero detuning to
the transitions from $|F=1,m_F=0\rangle$ to $|F'=0, m_{F'}=0\rangle$
and $|F=1,m_F=0\rangle$ to $|F'=1, m_{F'}=0\rangle$, respectively.
The probabilities for the generation of $\sigma^+$ and $\sigma^-$
photons are equal for three values of the detuning. They now deviate from a symmetrical picture with respect to the atomic resonances (grey vertical lines)
since the transition amplitudes via both excited states interfere.
(b) Time evolution of the system at $\Delta_{ca}/2\pi=63.2$\,MHz:
The probability density of the photon emission has a different shape
for the emission of a $\sigma^-$ photon (black) and $\sigma^+$
photon (grey). The final population inversion of the atomic state
differs from 1 because of losses due to spontaneous emission to
$|F=2\rangle$ and $|F=1,m_F=0\rangle$. This leads to losses in the
system which are about 10$\%$.} \label{fig:3-real}
\end{figure*}

In Fig.\,\ref{fig:3-real}\,(a) the calculated emission probability
is shown as a function of the cavity-atom detuning $\Delta_{ca}$,
with the black lines again showing emission probabilities for
$\sigma^-$ photons and grey lines for $\sigma^+$ photons. Compared
to Fig.\,\ref{fig:2-ideal}\,(b), the symmetry around
$\omega_c=\omega_{0e}$ is lost. At this frequency the probabilities
for $\sigma^+$ and $\sigma^-$ photon emissions differ from one
another although the cavity is in resonance with an atomic
transition. This can be qualitatively understood because the
influence of the $F'=1$ state becomes larger the closer the virtual
exited level is to this state. For a cavity-atom detuning of
$\Delta_{ca}=0$, the virtual excited level is between $F'=0$ and
$F'=1$ for the $\sigma^+$ process, but below the $F'=0$ level for
$\sigma^-$. Moreover, the detuning of the virtual level with respect
to the atomic bare states determines the sign of the transition
amplitudes. Therefore the two possible path of the Raman transition
(via $F'=0$ and $F'=1$) interfere either constructively or
destructively.

Although the former symmetry is lost, three cavity-atom detunings
$\Delta_{ca}$ are found where the efficiencies for $\sigma^+$ and
$\sigma^-$ photon production are equal. One is almost half-way
between the two atomic resonances, and the other two are close to
the $F=1$ to $F'=0$ and $F=1$ to $F'=1$ resonances. For the latter
two cavity frequencies, the probability for generating a photon
starting from the wrong initial state is below $7\%$
(Fig.\,\ref{fig:3-real}\,(a), dashed lines). This is an upper limit
for the probability of generating a second photon after a first
emission, if the process is started from the right initial state.

For instance, at $\Delta_{ca}/2\pi=63.2$\,MHz, the equal probabilities for $\sigma^+$ and $\sigma^-$ photon emission reach a maximum of $74\%$.
The time evolution of the system for this detuning is shown in Fig.\,\ref{fig:3-real}\,(b). The atom is exposed to the same pump pulse (shape and amplitude) for both directions of the Raman process. Although the probabilities for $\sigma^+$ and $\sigma^-$ photon emission are equal, the envelopes of
the emitted photons differ, as can be seen in the
probability-density plot. In fact, the virtual levels for the two
transitions are not at the same energy and their transition
amplitudes have different values. To produce identical photons, one
could eventually compensate for this by choosing suitably shaped
pump pulses for $\sigma^+$ and $\sigma^-$ processes. This is,
however, beyond the scope of this article. Differences in the two
processes can also be seen in the population transfer from the
initial to the final state, see Fig.\,\ref{fig:3-real}\,(b). It is
more successful for $\sigma^+$ photons, which indicates that the
losses to non-coupled states are higher when a $\sigma^-$ photon is
generated. Note that these losses never exceed $15\%$. Furthermore,
from the low probability of the wrong transition to take place, we
conclude that the starting conditions for a photon of opposite
polarisation are always met once a first photon has been emitted.
Therefore generating a sequence of photons of alternating
polarisation seems feasible.

\begin{figure}
\centerline{\includegraphics[width=\columnwidth]{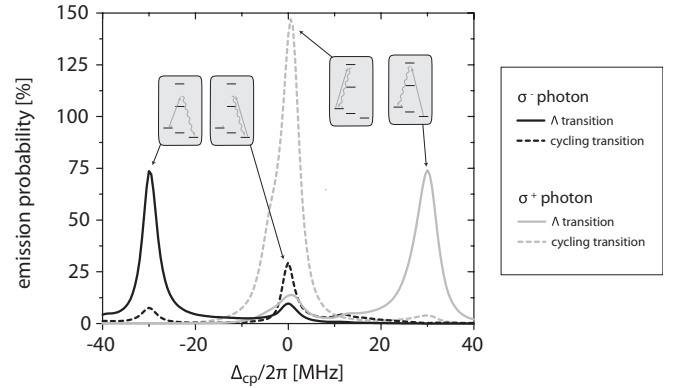}}
\caption{Dependency of the photon emission probability on the
frequency difference of pump laser and cavity. The cavity-atom
detuning is set to $\Delta_{ca}/2\pi = 63.2$\,MHz. The Raman
resonance of laser and cavity at $\pm 2\Delta_B$ leads to sharp
peaks in the photon emission probability for $\Lambda$ transitions
while cycling processes are only visible around $\Delta_{cp}=0$.}
\label{fig:4}
\end{figure}

We have seen that the production of a photon when the atom is
initially in the wrong Zeeman state is unlikely. We now show that
one can address either the $|{\scriptstyle -}\rangle$ to
$|{\scriptstyle +}\rangle$ or the $|{\scriptstyle +}\rangle$ to
$|{\scriptstyle -}\rangle$ transition by choosing the appropriate
pump laser frequencies. For a fixed cavity-atom detuning of
$\Delta_{ca}/2\pi=63.2$\,MHz, Fig.\,(\ref{fig:4}) shows the
calculated emission probabilities as a function of the cavity-pump
detuning, $\Delta_{cp}$. Again the pump laser Rabi frequency follows
a pulse amplitude as shown in Fig.\,\ref{fig:3-real}(b).
As expected, we find maxima in the photon emission probability
whenever Raman resonance conditions are met. Starting from state
$|{\scriptstyle -}\rangle$, the emission probability for $\sigma^-$
photons shows a maximum at a pump laser detuning of
$\Delta_{cp}=-30$\,MHz, and similarly starting from state
$|{\scriptstyle +}\rangle$ we find a maximum in the emission
probability of $\sigma^+$ photons for $\Delta_{cp}=+30$\,MHz. The
emission probability amounts to $74\%$, as discussed before. Note
that the Rayleigh scattering peak at $\Delta_{cp}=0$ dominates the
spectrum. It exceeds 100\% emission probability, since the pump
laser hits the cavity resonance and the atom undergoes a cycling
transition. For this reason, more than one photon per pulse can be
emitted. These cycling transitions are more pronounced for
$\sigma^+$ photons, since here the virtual excited level of the
Raman transition is closer to a real atomic level. To guarantee
single-photon emission in our scheme, transitions where two photons
are possibly emitted have to be avoided. Since the width of the
resonances depends on the cavity decay rate 2$\kappa$, the magnetic
field has to be chosen high enough to ensure that the separation of
the transition lines significantly exceeds 2$\kappa$. For the Zeeman
splitting considered here the Raman resonances are well resolved and
the scheme is not disturbed by cycling transitions, as can be seen
in Fig.\,(\ref{fig:4}).

\section{Conclusion}
The calculations discussed in this paper are based on the parameters
of an atom-cavity system that has been studied experimentally in our
group \cite{Hennrich00,Kuhn02,Hennrich03,Legero04}. In these
experiments, $^{85}$Rb was used and the Raman transitions were
driven between two different hyperfine ground states. The Zeeman
structure of the levels was not important. Moreover, a repumping
laser was necessary to re-establish the starting conditions after
each photon emission. The scheme proposed here could easily be
implemented if $^{87}$Rb atoms were used and a magnetic field was
applied. With the pump frequency being switched from one pulse to
the next in a way that the Raman transition is either driven from
$m_F =+1$ to $m_F =-1$ or vice versa, a stream of photons with
alternating polarisation is expected. No time-consuming repumping
will be needed, so that the photon-emission rate increases. The
simulations show that equal efficiencies can be obtained for the
production of $\sigma^+$ and $\sigma^-$ photons when an appropriate
cavity-atom detuning is used. The efficiency is 74$\%$ for a cavity
resonance close to the $F=1$ to $F'=1$ transition frequency. The
losses to other states are small, which guarantees that after a
first emission the atom is well prepared to produce a subsequent
photon. The realisation of this scheme seems promising. Since the
produced photons have a well defined polarisation, they can be
directed along arbitrary paths. This would be very convenient for
quantum communication and all-optical quantum information processing
in photonic networks \cite{Knill01}.

\section*{Acknowledgement}
This work was supported by the Deutsche Forschungsgemeinschaft (SFB
631, research unit 635) and the European Union [IST (QGATES,
SCALA) and IHP (CONQUEST) programs].

\end{document}